\documentclass[twocolumn,twoside,slac_two]{revtex4}
\usepackage{graphicx}
\usepackage{fancyhdr}
\newcommand{\megp}{$\mu^+\rightarrow\mathrm{e}^+\gamma\ $}
\newcommand{\meg}{$\mu\rightarrow\mathrm{e}\gamma\ $}

\newcommand{\radmup}{$\mu^+\rightarrow\mathrm{e}^+\nu_{\mathrm{e}}\bar{\nu}_{\mu}\gamma\ $}

\newcommand{\michelp}{$\mu^+\rightarrow\mathrm{e}^+\nu_{\mathrm{e}}\bar{\nu}_{\mu}\ $}
\newcommand{\Deg}{\(^{\circ} \) \kern-.4em \ }

\begin{document}

\title{Lepton Flavour Violating Muon Decay at MEG}

%

\author{H. Nishiguchi}
\affiliation{The University of Tokyo, 
             7-3-1 Hongo, Bunkyo 113-0033, Japan
\footnote{Present Address: 
          Institute of Particle and Nuclear Studies (IPNS), 
          High Energy Accelerator Research Organization (KEK),
          1-1 Oho, Tsukuba 305-0801, Japan}}

\begin{abstract}
  The MEG Experiment searches for a lepton flavour violating decay,
  \megp , with a branching-ratio sensitivity of $10^{-13}$ in order to 
  explore the parameter region predicted by many theoretical models
  beyond the Standard Model.
  Detector construction and the Engineering Run were completed in 2007,
  and the first Physics Run will be carried out in 2008.
  In this paper, the prospects of MEG Physics Run in 2008 is described 
  in addition to the experimental overview.
\end{abstract}

\maketitle

\thispagestyle{fancy}


\section{Introduction}
  The Standard Model of elementary particle physics
  is one of the greatest successes of modern science.
  Based on the principles of gauge symmetries
  and spontaneous symmetry breaking,
  everything had been consistently described 
  until experimental evidence for neutrino oscillation was shown 
  by SuperKamiokande for the first time.

  Now, {\bf Lepton Flavour Violation (LFV)} among charged leptons,
  which has never been observed while the quark mixing and the neutrino 
  oscillations have been experimentally confirmed,
  is attracting a great deal of attention, since its observation is highly 
  expected by many of well motivated theories beyond 
  the Standard Model\cite{motivation,muonreview}.
  Additionally, it would be a clear evidence of existence 
  of new physics beyond the Standard Model
  because it is strongly suppressed in the Standard Model.
  Even if we assume a finite neutrino mass within the Standard Model,
  \megp could occur with a negligible rate, $\approx 10^{-50}$.
  In consequence, recent reviews on flavour physics 
  (see Ref.\cite{motivation} for example) thus
  indicate high expectations for the next leading \megp search experiment,
  {\bf MEG} \cite{MEG99}, which is just starting
  the physics-data taking in 2008.\\

  The ambitious goal of the MEG experiment is
  to search for a \megp decay with an improved sensitivity 
  by at least two orders of magnitude over the current best limit of 
  $\mathcal{B}(\mu^+\rightarrow\mathrm{e}^+\gamma)
  <1.2\times10^{-11}$(90\%C.L.) \cite{MEGA}. 
  It is predicted that \megp is naturally causable 
  with a branching ratio just below the current 
  upper limit,$10^{-11}\sim10^{-14}$, by the leading theories for physics 
  beyond the standard model, {\it eg.} the Supersymmetric theories of Grand 
  Unification or Supersymmetric Standard Model with the 
  {\it seesaw} mechanism (see Ref.\cite{muonreview} for a review).

\begin{figure}[h]
      \begin{center}
        \includegraphics[height=3.4cm]
          {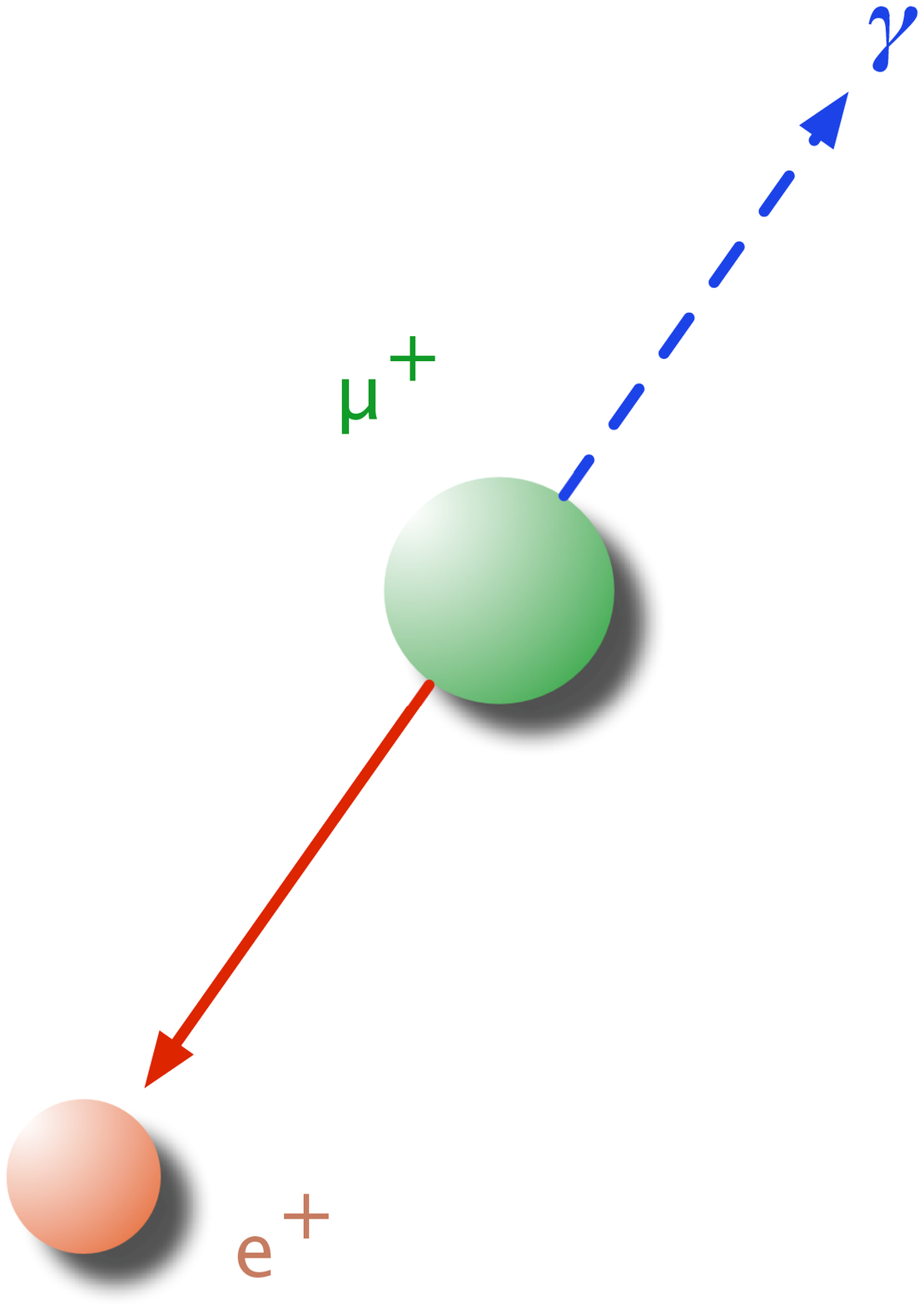} 
        \includegraphics[height=3.4cm]
          {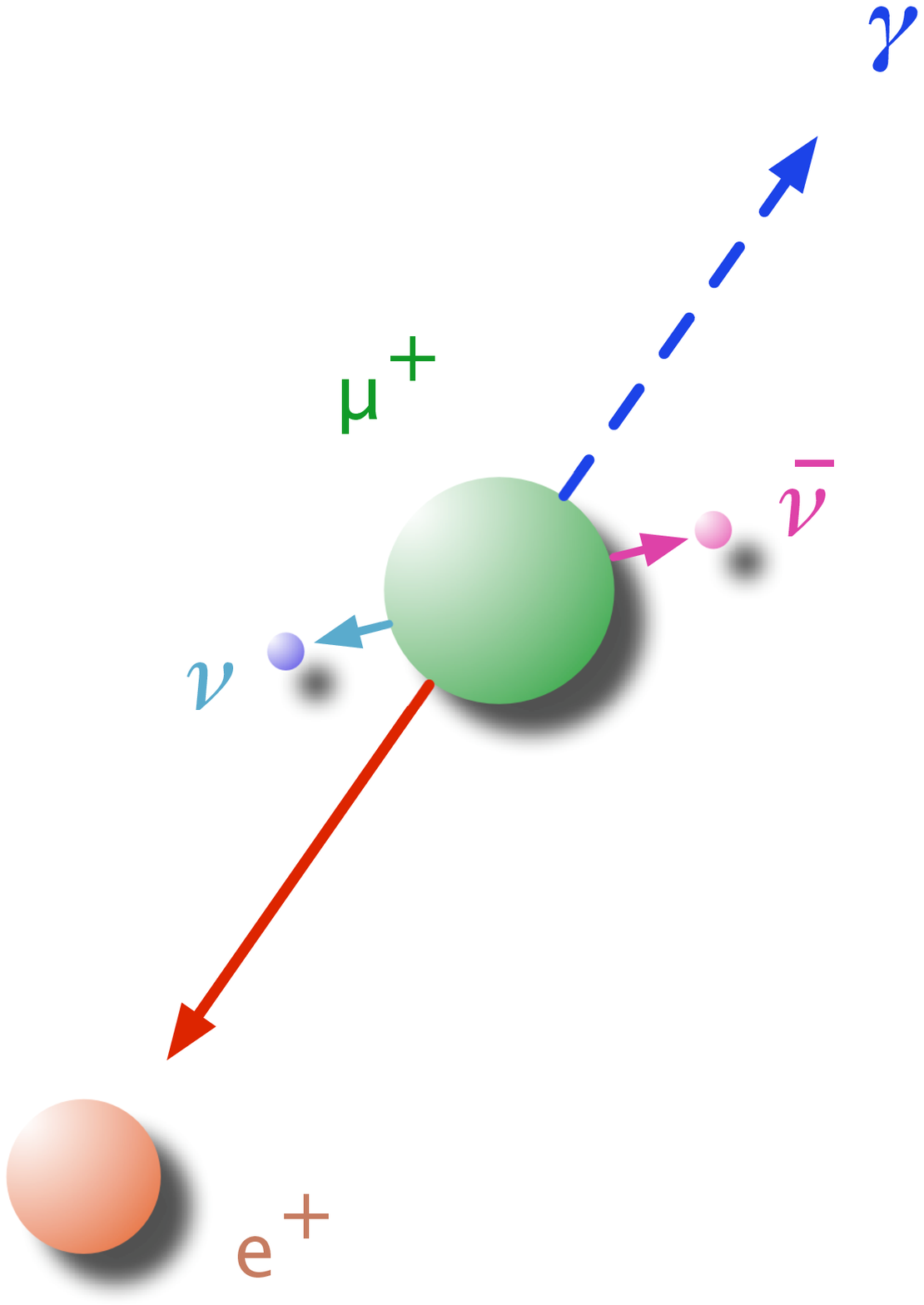} 
        \includegraphics[height=3.4cm]
          {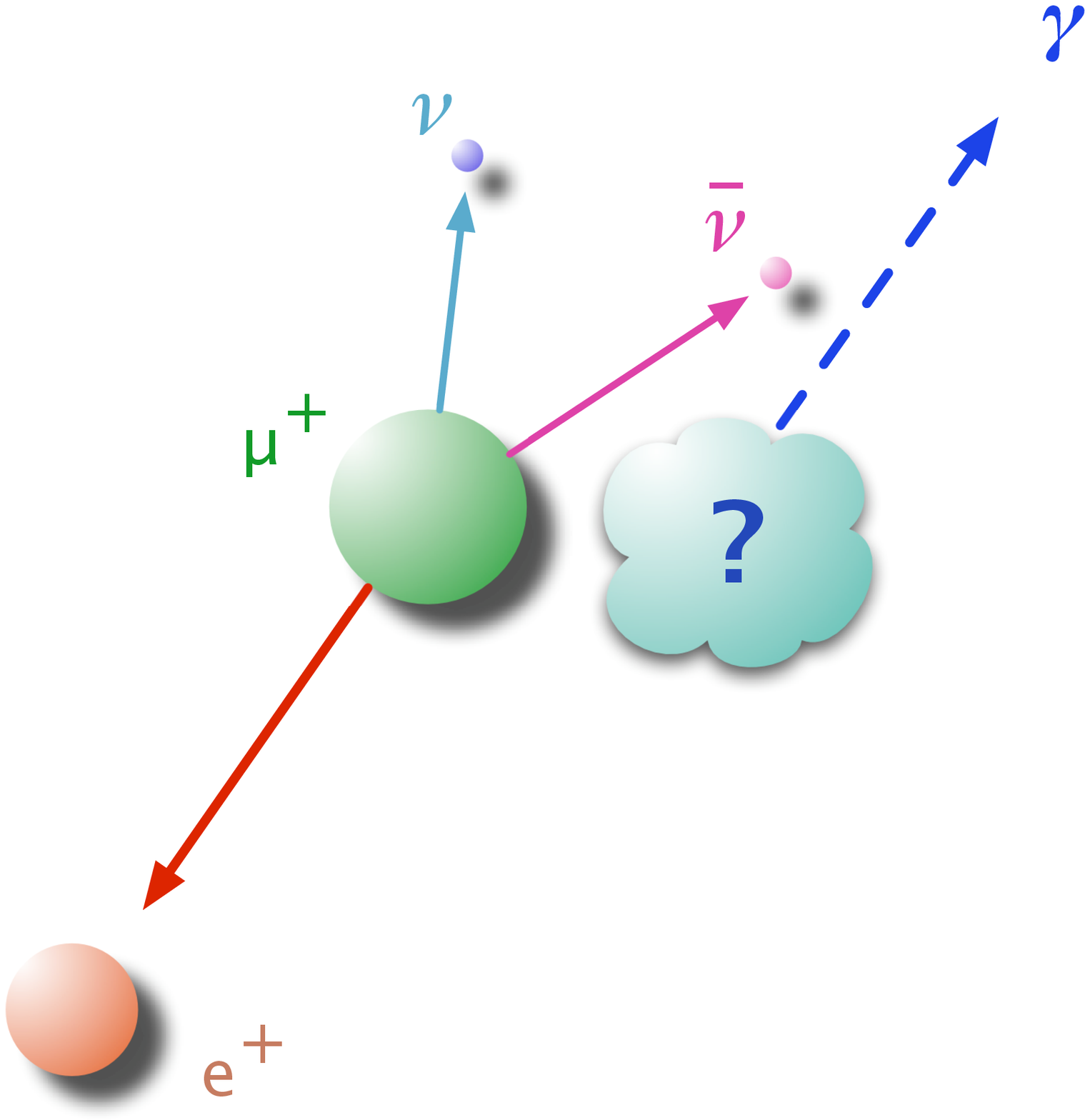} 
        \caption{Schematic views of \megp event signature and backgrounds;
                        (Left) \megp Signal, (Centre) Physics Background, (Right)
                        Accidental Background
                       \label{signal_bg}}
      \end{center}
    \end{figure}

  The signal of \megp decay is very simple and 
  is characterized by a 2-body final state of 
  a positron and $\gamma$-ray pair emitted in opposite directions 
  with the same energy, 52.8MeV, which corresponds to half the muon mass.
  There are two major backgrounds in the search for \megp.
  One is a physics (prompt) background from a radiative muon decay, \radmup,
  when the positron and the $\gamma$-ray are emitted back-to-back 
  with the two neutrinos carrying off tiny energy. 
  The other background is accidental coincidence of a positron
  from a normal Michel decay, \michelp, with a high energy random photon.
  The source of high energy $\gamma$ ray is either a radiative decay
  \radmup, annihilation-in-flight or external bremsstrahlung of a positron.
  Both are schematically shown in Figure \ref{signal_bg}
  in addition to the signal kinematics.
  The background is primarily dominated by accidental coincidence.
  Suppressing such an accidental overlap holds the key 
  for leading MEG to a successful conclusion.
  
  The excellent sensitivity of the MEG experiment is enabled 
  by three key elements:
  (1) the world's most intense DC muon beam provided at the Paul Scherrer Institute (PSI);
  (2) an innovative liquid xenon scintillation $\gamma$-ray detector 
        \cite{Xenon};
  (3) a specially designed positron spectrometer with a highly graded magnetic field \cite{COBRA}.
        
  The beam line and the detector construction
  have been completed in summer 2007, and 
  Beam- and Detector-Engineering run has been carried out right after that.
  In this engineering run, 
  all the detector-calibration procedures were established.
  In addition to the calibration, this engineering run provided 
  a lot of information, {\it eg.}
  detector performances, expected number of backgrounds, and what we have to
  maintain until the first physics data-taking
  in order to gain an experimental sensitivity as high as possible.

\section{Beam and Detector}
\begin{figure*}[htb]
    \centering
    \includegraphics[width=144mm]{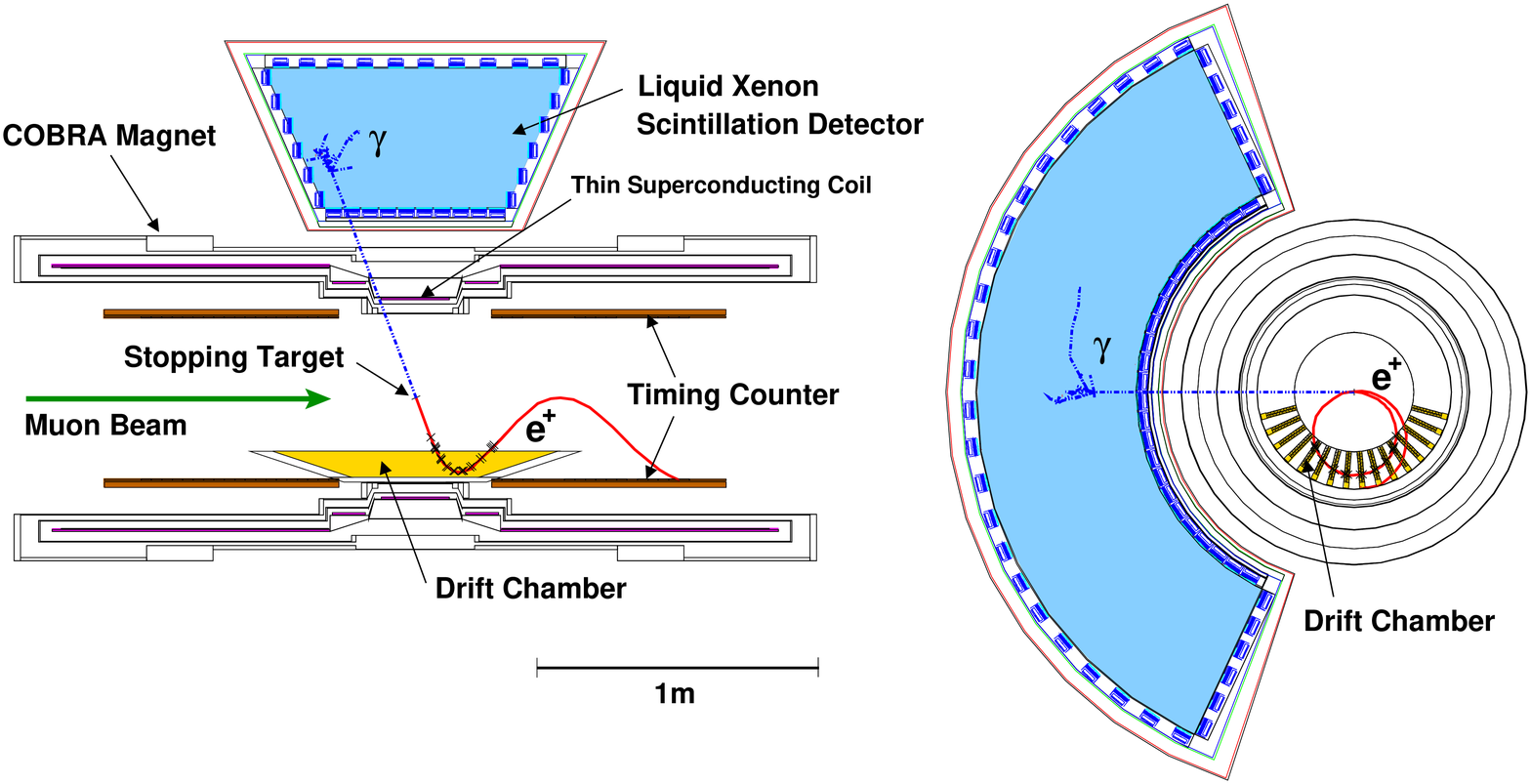}
    \caption{A schematic view of the MEG apparatus \label{apparatus}}
  \end{figure*}

  In order to fulfill the ambitious goal,
  the MEG experiment is designed carefully.
  The MEG detector apparatus consists of the muon beam transport system 
  and the detector system;
  the Photon Detector and the Positron Spectrometer.
  A schematic view of the MEG apparatus is shown in Figure \ref{apparatus}.\\

  A DC muon beam is the best tool to search for \megp
  since experimental sensitivity is mainly limited by accidental
  overlap of background events.
  The 590 MeV proton cyclotron at the Paul Scherrer Institute (PSI) 
  delivers up to 2.2 mA proton beam,
  which is the world record for such proton cyclotrons at present (2008).
  This megawatt accelerator has played a role of {\it progenitor} of the
  most intense DC pion and muon beams and
  made it possible to measure the
  rare decays and the search for ``classical'' forbidden decay modes,
  {\it eg.} \meg .
  The MEG experiment employs this most intense DC muon beam.\\
  
  The momentum and direction of positrons are measured precisely
  by a Positron Spectrometer, which consists of a superconducting solenoidal
  magnet specially designed to form a highly graded field,
  an ultimate low-mass drift chamber system, and a precise time
  measuring counter system \cite{COBRA}.
  The  Positron Spectrometer has to satisfy several requirements.
  First, the spectrometer must cope in a stable way with a very high muon rate 
  up to $3\times10^7$ s$^{-1}$.
  Second, a very-low-mass tracker is required since the momentum resolution
  is limited primarily by multiple Coulomb scattering.
  Furthermore, it is also important to minimize the amount of material
  from the point of view of background suppression in the photon detector.
  Additionally, excellent bidirectional spacial resolution of the tracker
  is necessary for both the transverse and longitudinal directions.
  Finally, excellent timing resolution is also necessary in order to suppress
  accidental overlap of events.

  In order to attain such requirements, we adopted a specially designed solenoidal
  magnet with a highly graded field.
  The MEG solenoidal magnet is designed to change its radius
  between the centre and the outside.
  This provides a highly graded magnetic field
  (1.27 T at $z=0$ and decreasing as
  $|z|$ increases, 0.49 T at $z=1.25$ m, where $z$ is the coordinate along 
  the beam axis) and allows to solve the problems inevitable in a normal 
  uniform solenoidal field.
  In a uniform solenoidal field, positrons that emitted close to 90\Deg 
  undergo many turns in the tracker volume.
  However, the MEG solenoidal magnetic field can sweep such positrons 
  out of the fiducial tracking volume quickly.
  In addition, this special magnetic field has yet another advantage.
  In this specially designed field, positrons with the same absolute momenta
  follow trajectories with a constant projected bending radius
  independent of the emission angles
  while in a uniform solenoidal field
  the bending radius depends on the emission angle.
  This allows us to discriminate sharply high momentum signal positrons
  from the tremendous Michel positron background originating from the 
  muon-stopping target.
  The Positron Spectrometer therefore does not need to measure 
  the positron trajectory in the small radius region.
  In other words, the drift chambers can be sensitive
  only to higher momentum positrons and blind to most of the Michel positrons
  that can cause accidental coincidences.
  Thanks to this benefit, this spectrometer can cope within
  such a highly-irradiated environment.\\
  
  While all positrons are confined by the solenoid,
  the $\gamma$ ray pass through the thin superconducting coil
  of the spectrometer with $\approx$80\% 
  transmission probability,
  and are detected by an innovative liquid-xenon photon detector
  \cite{Xenon}.
  Scintillation light emitted inside liquid xenon are viewed
  from all sides by photo-multiplier tubes (PMT)
  that are immersed in liquid xenon in order to maximize 
  direct light collection.
  Liquid-xenon scintillator has very high 
  light yield ($\approx$75\% of NaI crystal) 
  and fast response, which are the most essential ingredients 
  for precise energy and timing resolutions required for this experiment.
  A scintillation pulse from xenon is very fast and has a short tail,
  thereby minimizing the pile-up problem.
  Distributions of the PMT outputs enable a measurement
  of the $\gamma$-ray incident position with a few mm accuracy.

  Absorption of scintillation light by impurities
  inside liquid xenon, especially water and oxygen,
  could significantly degrade the detector performance,
  although there is no absorption by liquid xenon itself.
  In order to solve the absorption issue,
  a purification system that circulates and
  purifies xenon gas was developed \cite{gas_purification}.
  Various studies were carried out using a
  100 liter prototype detector with 238 PMTs 
  in order to gain practical experiences in operating
  such a new device and to prove its excellent performance.
  The prototype detector was tested by using $\gamma$ rays
  from laser Compton scattering at National Institute of
  Advanced Industrial Science and Technology (AIST) in Tsukuba, Japan.
  Gamma rays with the Compton edge energy of 10, 20 and 40 MeV
  were generated via backward scattering of laser photons
  by 800 MeV electron beam in the storage ring of AIST.
  Another test was carried out at PSI by using the
  pion charge exchange reaction, 
  $\pi^{-}p\rightarrow\pi^{0}n$,
  which provides two $\gamma$ rays from the $\pi^{0}$ decay.
  By tagging back-to-back $\gamma$ rays,
  monochromatic $\gamma$ rays of 55 MeV and 83 MeV are selected.
  The energy resolution of 2 \%, the timing resolution of 65 ps,
  and the position resolution of $\approx$4 mm depending on the 
  incident position with respect to the PMT positions were obtained
  by these beam tests.
  In addition to the performance estimation,
  the purification method was established.

  Based on the prototype works,
  the final MEG liquid-xenon photon detector,
  which is filled with $\approx$900 litres of liquid xenon
  incorporating 846 PMTs, was built.
  In order to speed up the purification process for the final detector,
  a liquid-phase purification system that uses a cryogenic
  centrifugal fluid pump capable to flow 100 liter of
  liquid xenon per hour was developed \cite{liq_purification}.
  Currently(2008), this is the world's largest liquid-xenon photon detector.\\

  All the signals, from PMTs and drift chambers, are individually recorded
  as digitized waveform by a custom chip called Domino Ring Sampler (DRS)
  \cite{DRS}.
  The PMT signals are digitized at 1.6 GHz sampling speed
  to obtain a timing resolution of 50 ps by bin interpolation,
  and the drift chamber signals are digitized at 500 MHz
  in order to compensate wide drift time distributions.
  Recording all the waveform may cause difficulties concerning
  data size, data-acquisition (DAQ) flow speed {\it etc.},
  however the rewards outweigh the works and difficulties,
  because waveform digitizing of all channels gives us 
  an excellent handle to identify the pile-up event
  and to suppress noise that can worsen detector resolutions.
  
\section{Run 2007 (Engineering Run)}
  In summer 2007, construction of all detector components 
  was completed.
  Then we immediately started the detector operation in a phased manner.

  We started several studies for the liquid-xenon photon detector,
  liquefaction test, liquid xenon transferring test, 
  long term monitoring, stability checks, PMT gain calibrations, 
  and liquid-xenon purification {\it etc}.
  In parallel with this, positron-spectrometer conditioning was performed,
  drift-chamber gas control test, high-voltage conditioning,
  wire alignment by using cosmic rays, position-measurement calibration,
  relative gain calibration {\it etc}.
  After such fundamental studies and conditioning,
  the muon-beam commissioning was performed with the final detector apparatus.
  Final focusing of the muon beam, beam profile measurement,
  and muon rate measurement, were carried out.

\begin{figure}[htb]
    \includegraphics[width=60mm]{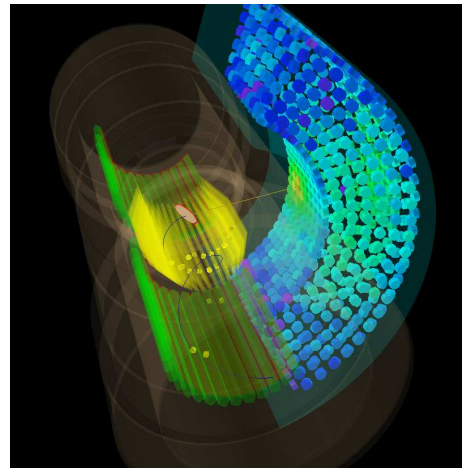}
    \caption{An example of event display of Engineering-Run 2007 \label{event}}
  \end{figure}

  After the muon-beam commissioning was completed, 
  we started Engineering Run in October 2007.
  Figure \ref{event} shows a typical example
  of accidental background events in the Engineering-Run.
  We successfully ran the whole program of the Engineering Run.
  All the detector components were operated over three months,
  trigger and DAQ electronics were integrated and data-taking worked
  at expected event rate, a full set of calibration has been performed.
  The physics data that has been taken at the Engineering Run 
  was analyzed in the winter shutdown 2007-2008,
  and its results gave a certain feedback
  to the detector maintenance and also the offline-analysis development.
  One of the most important analysis of Engineering Run 2007
  are to complete the reconstruction algorithm for all the detector,
  to evaluate the detector performances,
  and to estimate the feasible sensitivity of MEG.\\

  By analyzing the data of engineering-run 2007,
  we could evaluate all the detector performances and
  verify the quality of Monte Calro (MC) simulation for the 
  MEG detector apparatus.
  Unfortunately, the obtained performances were
  little worse than the design value
  due to several unsatisfactory conditioning of detectors,
  {\it eg.}
  approximately 27 \% of dead channel of the
  drift-chamber system degraded the spectrometer resolutions,
  still remained impurity of liquid xenon 
  deteriorated the photon-detector resolutions,
  and badly-fabricated internal clock circuit of DRS
  provided poor timing resolutions for all sub-detectors.
  Although such bad performances were obtained,
  we could figure out the sources of deterioration
  and reproduce phenomenon by the MC incorporating artificially-degraded
  detector descriptions.

  Table~\ref{performances} summarizes the obtained performances
  by the Engineering-Run 2007 and the expected performances
  for the MEG Physics-Run 2008.
  Expected values are obtained by assuming that
  all the deterioration clarified in 2007 
  could be fixed by the maintenance works during winter shutdown
  of 2007-2008.
  All resolutions are converted to Full-Width-at-Half-Maximum (FWHM).

  \begin{table}[h]
  \begin{center}
  \caption{Detector Performances, Obtained(2007) and Expected(2008)
           \label{performances}}
  \begin{tabular}{l|c|c}
    \hline 
    \hline {Quantity} & {{~}Run 2007{~}} & {{~}Run 2008{~}} \\
    \hline 
    \hline 
      $\gamma$-Energy Resolution (\%)   & 6.5 & 5.0 \\
      $\gamma$-Timing Resolution (ns)   & 0.27 & 0.15 \\
      $\gamma$-Spatial Resolution (mm)  & 15 & 9.0 \\
      $\gamma$-Detection Efficiency (\%)   & $>$40  & $>$40 \\
    \hline
      e$^+$-Momentum Resolution (\%){~} & 2.1 & 1.1 \\
      e$^+$-Timing Resolution (ns)    & 0.12 & 0.12 \\
      e$^+$-Angular Resolution (mrad) & 17 & 17 \\
      e$^+$-Detection Efficiency (\%) & 39 & 65 \\
    \hline
    \hline 
  \end{tabular}
  \end{center}
  \end{table}

\section{Run 2008 (Physics Run)}

  On the basis of the result of the Engineering-Run 2007,
  we completed various maintenance on each sub-detectors 
  during the winter-shutdown term 2007-2008,
  and now we are carrying out the final detector conditioning.
  The MEG Physics Run is being planned to start in summer 2008.
  We here discuss the feasible sensitivity of the MEG experiment
  by employing obtained performances and expected improvements.

  We are planning to have 20 weeks of beam time in 2008.
  According to the PSI proton-accelerator operation procedure,
  20-weeks beam time is corresponding to
  8$\times 10^{6}$ sec.
  By employing these numbers with expected beam intensity,
  3$\times 10^{7}$ sec$^{-1}$, 
  the number of background event for the MEG physics run 2008
  is expected to be 0.4.

  Finally, let us evaluate the single event sensitivity
  and the feasible upper limit that will be determined 
  by physics run 2008.
  By assuming 65 \% of positron-detection efficiency,
  40 \% of $\gamma$-ray detection efficiency,
  70 \% of selection efficiency,
  3$\times 10^{7}$ s$^{-1}$ of muon-beam intensity,
  $T=8\times 10^{6}$ s of experiment-running time,
  and $\Omega/4\pi = 0.09$ of detector solid angle that is 
  calculated from the detector geometrical acceptance,
  the single event sensitivity for the MEG physics run 2008
  can be evaluated as
  $
    \mathcal{B}^{2008}_{\mathrm{S.E.S.}}(\mu^{+}\rightarrow\mathrm{e}^{+}\gamma)
     = 2.6 \times 10^{-13}.
  $
  In the case that no candidate is observed,
  a $2.6\times10^{-13}$ single event sensitivity
  implies the upper limit on
  $\mathcal{B}^{2008}
  (\mu^{+}\rightarrow\mathrm{e}^+\gamma)$
  at the 90 \% confidence level as
  $
     < 7.2 \times 10^{-13}
  $
  for the MEG physics run 2008.
  This sensitivity will be eventually improved
  down to $1\times10^{-13}$
  by the follow-up data-taking after next year.


\bigskip 

\end{document}